\documentclass[fleqn,usenatbib]{mnras}
\usepackage[utf8]{inputenc}
\usepackage[catalan, english]{babel}
\usepackage{graphicx}
\graphicspath{{./figures/}}
\usepackage{natbib}
\usepackage{appendix}
\usepackage{tabularx}
\usepackage{color}
\usepackage{pdflscape}
\usepackage{booktabs}
\usepackage{threeparttable}
\usepackage{longtable}
\usepackage{multicol}
\usepackage{multirow}
\usepackage[normalem]{ulem}
\usepackage{enumerate}
\usepackage{amsmath}
\usepackage{amssymb}
\usepackage{setspace} 
\usepackage{hyperref}

\hypersetup{%
    colorlinks=false,linktoc=all, pdfstartpage=1, pdfstartview=FitV, pdfencoding=auto,%
    breaklinks=true, pdfpagemode=UseNone, pageanchor=true,
    pdfpagemode=UseOutlines,%
    plainpages=false, bookmarksnumbered, bookmarksopen=true, bookmarksopenlevel=2,%
    hypertexnames=true, pdfhighlight=/O,
    urlcolor=webbrown, linkcolor=RoyalBlue, citecolor=webgreen,
}

\definecolor{halfgray}{gray}{0.55}
\definecolor{naviblue}{RGB}{0,0,102}
\definecolor{webbrown}{rgb}{.6,0,0}
\definecolor{RoyalBlue}{cmyk}{1, 0.50, 0, 0}
\definecolor{webgreen}{rgb}{0,.5,0}
\definecolor{Maroon}{cmyk}{0, 0.87, 0.68, 0.32}
\definecolor{Black}{cmyk}{0, 0, 0, 0}
\definecolor{myorange}{RGB}{239, 186,67}

\def\etal{et al.\ }

\newcommand{\lyb}{Ly\texorpdfstring{$\beta$}{beta}}
\newcommand{\lya}{Ly\texorpdfstring{$\alpha$}{alpha}}
\newcommand{\hmpc}{h$^{-1}$Mpc}
\newcommand{\kms}{km/s}

\begin{document}

\title{The cross-correlation of galaxies in absorption with the Lyman \texorpdfstring{$\alpha$}{alpha} forest}
\date{}
\author[Ignasi P\'erez-R\`afols \etal]
  {Ignasi ~P\'erez-R\`afols,$^{1,2,3}$\thanks{email: iprafols@gmail.com}, Matthew M. Pieri$^{3}$, Michael Blomqvist$^{3}$, Sean Morrison$^{3, 4}$,
  \newauthor Debopam Som$^{5}$, Andrei Cuceu$^{6, 7}$
  \\
$^{1}$Institut de Física d’Altes Energies, The Barcelona Institute of Science and Technology, Campus UAB, 08193 Bellaterra (Barcelona), \\Spain\\
$^{2}$Dept. Física Quàntica i Astrofísica, Institut de Ciències del Cosmos (ICCUB), Facultat de Física, Universitat de Barcelona (IEEC-UB), \\Martí i Franquès, 1, E08028 Barcelona, Spain\\
$^{3}$Aix Marseille Univ, CNRS, CNES, LAM, Marseille, France\\
$^{4}$Department of Astronomy, University of Illinois at Urbana-Champaign, Urbana, IL 61801, USA\\
$^{5}$Space Telescope Science Institute, 3700 San Martin Drive, Baltimore, MD 21218, USA\\
$^{6}$Center for Cosmology and Astro-Particle Physics, The Ohio State University, Columbus, Ohio 43210, USA\\
$^{7}$    Department of Astronomy, The Ohio State University, Columbus, Ohio 43210, USA\\
}

\maketitle

\begin{abstract}
    We present the first clustering measurement of Strong Blended Lyman $\alpha$ (SBLA) absorption systems by measuring their cross-correlation with the Lyman $\alpha$ forest. SBLAs are a new population of absorbers detected within the Lyman $\alpha$ forest. We find a bias of $2.329\pm0.057$, consistent with that of Damped Lyman $\alpha$ absorbers (DLAs). For DLAs, we recover a bias of $2.331\pm0.057$ larger than previously reported \citep{Perez-Rafols+2018a}. We also find a redshift space distortion parameter $\beta=0.417\pm0.010$, also consistent with the recovered value for DLAs ($\beta=0.416\pm0.010$). This is consistent with SBLA and DLA systems tracing different portions of the circumgalactic medium of a broadly common population of galaxies. Given these common clustering properties, we combined them to perform a cross-correlation of galaxies in absorption with the \lya{} forest. We find that the BAO scale uncertainty of this new measurement is $1.75\times$ that of \lya{} auto-correlation and $1.6\times$ that of the quasar cross-correlation with the \lya{} forest. We note that the current preferred metal contamination model for fitting the correlation functions with respect to the \lya{} forest is not realistic enough for SBLA systems, likely due to their status as high redshift precision sites of high metal enrichment. Mock spectra including SBLA systems and their associated metal absorption are required to understand this sample fully. We conclude that SBLAs have the potential to complement the standard \lya{} cosmological analyses in future surveys.
\end{abstract}



\section{Introduction}
The standard $\Lambda$CDM model for cosmology describes an expanding Universe. This expansion is presently accelerating. The origin of this acceleration is currently one of the greatest open questions of modern physics. The first evidence for this acceleration was seen by luminosity distances to type Ia supernovae \citep[SNe IA, ][]{Riess+1998, Perlmutter+1999}. Baryonic Acoustic Oscillations (BAO) in the matter correlation function support this evidence by measuring the expansion rate normalised to the sound horizon $r_{d}$. The comoving scale of the BAO peak $r_{d}$ acts as a "comoving standard ruler". It can be estimated, for example, using CMB anisotropies which, in turn, provide indirect evidence of this accelerated expansion \citep{Planck+2016, Planck+2020}. The latest constraints on cosmological parameters from BAO \citep{Alam+2021} are
consistent with those from SNe Ia \citep{Scolnic+2018, Jones+2019} and CMB measurements.

BAO measurements are performed on various redshifts using various tracers depending on the availability of those tracers. At redshifts $z\lesssim2$ the BAO scale is measured using discrete tracers such as galaxies \citep[e.g.][]{Alam+2017, Bautista+2018},
galaxy clusters \citep{Hong+2016}, and quasars \citep{Ata+2018}. At higher redshifts, the number density of discrete tracers decreases and becomes insufficient for high-precision clustering measurements. At these redshifts, BAO has been measured using the Lyman $\alpha$ (\lya{}) absorption  from the Intergalactic and Circumgalactic Media (IGM and CGM, respectively) observed in the so-called \lya{} forest. The BAO
scale was first measured in the \lya{} auto-correlation function \citep{Busca+2013, Slosar+2013, Kirkby+2013, Delubac+2015, Bautista+2017, deSainteAgathe+2019}, and then in the \lya{} $\times$ quasar cross-correlation
function \citep{Font-Ribera+2014, duMasdesBourboux+2017, Blomqvist+2019}.

A key potential source of systematic errors in the measurement of BAO in the \lya{} forest is its associated metal absorption. This absorption generates additional line-of-sight correlations associated with the ratio of restframe wavelengths for the transitions in question \citep{Slosar+2011, Pieri+2014, Bautista+2017, duMasdesBourboux+2020}. Indeed, it has further been argued that metal transitions can be analysed as a forest and used as large-scale structure tracers themselves \citep{Pieri2014, Blomqvist2018, duMasdesBourboux2019}. However, much of this metal absorption can be characterised (particularly in correlation functions that weight by line strength) as arising due to galaxies in absorption. Here we define `galaxies in absorption' to mean both the lines of sight passing through the galaxy itself and the regions around them, which they dominate gravitationally (known as the CGM and typically defined as within the virial radius of the hosting galaxy). Metal absorption either as a structure tracer or as a structure contaminant is closely connected to the presence and properties of galaxies in absorption. Hence attempts should be made to treat galaxies in absorption embedded in the \lya{} forest as distinct collapsed systems in order to measure their properties and their impact for measuring BAO. This is the subject of this work and its companion publication \citep{Morrison+Inprep}.

A variety of information is present in the broadband shape of the correlation function of the \lya{} forest beyond the measurement of BAO alone (e.g. \citealt{Cuceu+2021}). A limitation of this deeper exploitation is the fact that the \lya{} is a mixed tracer. The \lya{} forest traces both diffuse gas structures (in the IGM) and gas in the collapsed structures (in the CGM). As a result, the bias and redshift space distortions of the forest are hybrid values reflecting a mixed tracer and this degeneracy must be marginalised over \citep{Cuceu2022}. The separation of this mixed tracer into diffuse and collapsed components not only promises to allow for a cleaner measurement and fit of \lya{} forest correlation functions, but it also promises to deliver a distinct set of galaxies in absorption as a supplementary tracer. 

An additional challenge is presented by the fact that the broad absorption profile associated with Damped \lya{} (DLA) absorption systems (along with sub-DLAs) in the \lya{} leaves a measurable imprint on the correlation function \citep[e.g.]{Rogers+2018}. This can be somewhat mitigated by the masking and/or correction of known DLAs in the data, but not all DLAs are identified and their contribution must be taken into account in model fits to  correlation functions with respect to the \lya{} forest.

The goal of this paper is to take initial steps towards identifying this larger set of galaxies in absorption, mask them from the \lya{} forest sample and treat them as distinct tracers in their own right.

The clustering properties of DLAs were previously studied by \cite{Font-Ribera+2012} and \cite{Perez-Rafols+2018a,Perez-Rafols+2018b}, but no BAO measurement was reported since the number of DLAs was deemed to be too low. The DR16 DLA sample of \cite{Chabanier+2021} offers a much larger sample than can be used to measure BAO. These systems represent rare cases where very high column densities (($N_{\rm HI}>2\times10^{20}{\rm cm^{-2}}$) are present in the circumgalactic medium systems in the forest amoung systems that show self-shielded properties $N_{\rm HI} \gtrsim 10^{17}{\rm cm^{-2}}$). \cite{Pieri+2014} found that a larger sample of strong and blended \lya{} forest systems, with column densities too low for self-shielding to occur, are also typically tracing galaxies in absorption.
Updated studies, including more data, seem to support this picture (\citealt{Yang+2022}). \cite{Morrison+Inprep} explore this further and provide the sample of galaxies in absorption used in this work, which they have classed as `strong blended \lya{}' or SBLA systems. We explore the use of SBLAs and DLAs together as a new sample tracer of collapsed structures embedded in the diffuse \lya\ forest. 

This paper is organized as follows. In section~\ref{sec:data} we present the DLA and SBLA catalogues, and the \lya{} flux transmission field. These are used to measure the correlation functions as described in section~\ref{sec:xcf}. We present our results for the DLAs and the SBLAs in section~\ref{sec:results} and discuss its inclusion to the standard \lya{} BAO analysis in section~\ref{sec:joint_fit}. In section~\ref{sec:discussion} we discuss the implications of our work: on the measured biases, for cosmological mocks and on upcoming surveys. Finally, we summarize our findings and conclusions in section~\ref{sec:conclusions}.

Throughout this paper, we use version 4 of the python package \texttt{picca}\footnote{available at \url{https://github.com/igmhub/picca/releases/tag/v4}} \citep{picca} We use the package \texttt{vega}\footnote{Available at \url{https://github.com/andreicuceu/vega/tree/master/vega}} for modelling and fitting correlation functions.

\section{Data samples}\label{sec:data}
The study presented here uses different catalogues that are based mainly on the Data Release 16 (DR16) of the extended Baryon Oscillation Spectroscopy
Survey \citep[eBOSS,][]{Dawson+2016}, part of the Sloan Digital Sky Survey IV \citep[SDSS-IV,][]{Blanton+2017}.

\subsection{DLA sample}
We use the DLA sample from \cite{Chabanier+2021}\footnote{Available at \url{https://drive.google.com/drive/folders/1UaFHVwSNPpqkxTbcbR8mVRJ5BUR9KHzA}}. DLAs in this sample were identified using the convolutional neural network algorithm of \citep{Parks+2018}. \cite{Chabanier+2021} searched for DLAs only in quasars not identified as Broad Absorption Line Quasars and in the wavelength window $90.0 \leq \lambda \leq 134.6{\rm nm}$ in the quasar rest frame. 
We refer the reader to these two papers for a detailed description of the procedure. We note that this catalogue was used by \cite{duMasdesBourboux+2020} in their masking procedure (see section~\ref{sec:lya}).

In this catalogue, two estimates of the column density are given: one from the Convolutional Neural Network, \texttt{NHI\_CNN}, and one from a profile fit, \texttt{NHI\_FIT}. Results from \cite{Chabanier+2021} indicate that the latter is less biased (and thus preferable), but it is only computed for DLA candidates in the rest-frame range $104.0 \leq \lambda_{\rm RF} \leq 121.6{\rm nm}$ and $\log N_{\rm HI} < 22$. We keep \texttt{NHI\_FIT} as our $N_{\rm HI}$ estimate, whenever it is available, and we use \texttt{NHI\_CNN} when it's not. 

The catalogue contains a total of 117,458 DLA candidates with $2 \leq z_{\rm DLA} \leq 5.5$ and $19.7 \leq \log N_{\rm HI} \leq 22$. From this initial sample, we construct a pure sample of DLAs (with purity $>0.9$) by restricting to candidates with $z_{\rm DLA} < 3.2$ and $20.5 < \log N_{\rm HI} < 21.5$ \citep{Chabanier+2021}. This reduces the number of DLAs candidates to 29,368. Of these, 3 DLAs are identified in sight lines with \texttt{THING\_ID} $>$ 0. This normally means some problem occurred during the reduction process and so we discard these DLAs. The final number of DLAs used is 29,365. 


\subsection{SBLA sample}
As described in \cite{Morrison+Inprep}, SBLAs are defined as systems with true flux transmission less than 25\% over a $\sim138$\kms{} region of the forest. Since \lya\ forest lines are always intrinsically narrower (excluding DLAs, which we mask), this only occurs when extended absorption structures are present. \cite{Pieri+2010}, \cite{Pieri+2014} and \cite{Morrison+Inprep} set out the argument for why SBLAs appear to be galaxies in absorption, in particular using Lyman break galaxy samples and physically properties absorbers associated with SBLAs.

We construct an extended version of a fixed purity sample of prioritising higher completeness and purity of 30\% using a criterion for purity described in \cite{Morrison+Inprep}. Here we summarize the procedure used for the detection but refer the reader to \cite{Morrison+Inprep} for more details including a characterisation of the sample. 

SBLAs are selected from quasar spectra with some basic pre-selection criteria applied. These are extracted from the DR16Q quasar catalogue \citep{Lyke+2020} following the procedure indicated in \cite{Lee+2013} in order to compute PCA continua of the \lya\ forest.
We restrict to quasars with redshift $z\geq2.15$, without detection of Broad Absorption Lines, with a median signal-to-noise $S/N\geq0.2{\rm pixel}^{-1}$ in the \lya{} forest and $S/N\geq0.5{\rm pixel}^{-1}$ over the region $126.8-138.9{\rm nm}$, and with less than 20\% of the pixels flagged to be unreliable in the rest-frame wavelength range $121.6-160.0{\rm nm}$. The first cut ensures that there are enough \lya{} pixels to fit the PCA
continua. The rest of the cuts ensure that these PCA
continua are reliable.

From this subset of quasar spectra, we initially pre-select those with a transmitted flux  $-0.05\leq F< 0.25$ over bins of $\sim138$\kms{} in the rest-frame wavelength range $104.1-118.5{\rm nm}$. In \cite{Morrison+Inprep} they restrict the redshift of the SBLA candidates to be $2.4 < z_{\rm abs} <3.1$, but here we extend this range to be $1.94 < z_{\rm abs} < 5.45$. The SBLA sample is built by finding a maximum observed (noise-in) flux limit that returns a true (noise-free) flux of $F< 0.25$ with 30\% purity. This maximum observed flux is dependent on the S/N of each \lya{} forest and its functional dependence on S/N was derived using \lya{} forest mock spectra of SDSS DR11  dataset \citep{Bautista+2015}. Our final sample has a total of 742,832 SBLAs. The need for this S/N dependent flux maximum is driven by the fact that the flux transmission probability distribution is skewed towards weaker absorption across our idealised $F<0.25$ limit, which in turn drives the purity.

The mock spectra are of acceptable quality for the studying purity since all that requires is an acceptable representation of the diffuse \lya{} forest. Understanding completeness is more challenging since it requires a realistic SBLA population, which is not present in these mocks. Indeed, one of the goals of this initial study is to stress the importance of these mocks (and thus motivate their creation). For the purposes of this work, we prioritise completeness over purity in order to deal with this uncertainty. In doing this we prioritise the measurement BAO over reliable measurements of SBLA bias. A measurement of SBLA bias from a purer sample will be presented in \cite{Morrison+Inprep}. Nevertheless, our completeness will not be 100\% and the role of missing SBLAs must be considered.  The most obvious impact is  the loss of potential structure tracers. A further effect is that missed SBLAs are not masked (see section~\ref{sec:lya}). Both effects combined impact our goal of separating collapsed and diffuse parts of the \lya{} forest in search of optimal correlation function signal and simpler correlation function fitting.


\subsection{\protect{\lya{}} sample}\label{sec:lya}
DLAs and SBLAs are cross-correlated with the flux transmission field extracted from the \lya{} forest, $\delta_{q}\left(\lambda\right)$. We use the two catalogues from \cite{duMasdesBourboux+2020}\footnote{Available at \url{https://dr16.sdss.org/sas/dr16/eboss/lya/}}, corresponding to the transmission field of the ``\lya{} region", $\lambda_{\rm RF} \in [104, 120]$ nm and the ``\lyb{} region", $\lambda_{\rm RF} \in [92, 102]$ nm containing the transmission field from 210,005, and 69,656 sightlines respectively.

The extraction of the flux transmission field is explained in detail in \cite{duMasdesBourboux+2020}, but we summarize it here. The mean transmission field  at the observed wavelength, $\lambda$, is obtained from the ratio of the observed flux, $f_{q}\left(\lambda\right)$, and the mean expected flux, $\overline{F}(z)C_{q}\left(\lambda\right)$:
\begin{equation}
    \delta_q\left(\lambda\right)=\frac{f_{q}\left(\lambda\right)}{\overline{F}(z)C_{q}\left(\lambda\right)} - 1 ~,
\end{equation}
where $\overline{F}(z)$ is the mean transmission and $C_{q}\left(\lambda\right)$ the unabsorbed quasar continuum. 

The unabsorbed quasar continuum is estimated from the spectra themselves, distorting the field. To correct this, we apply the following transformation
\begin{equation}
    \label{eq:distortion_correction}
    \delta_q\left(\lambda_{i}\right) \rightarrow \sum_{j} \eta^{q}_{ij}\delta_q\left(\lambda_{j}\right) ~,
\end{equation}
where
\begin{equation}
    \eta^{q}_{ij} = \delta^{K}_{ij} - \frac{w_{j}}{\sum_{k}w_{k}} - \frac{w_{j}\left(\Lambda_{i}-\overline{\Lambda_{q}}\right)\left(\Lambda_{j}-\overline{\Lambda_{q}}\right)}{\sum_{k}w_{k}\left(\Lambda_{k}-\overline{\Lambda_{q}}\right)^2} ~,
\end{equation}
and $\overline{\Lambda_{q}}$ is the mean of $\Lambda=log\lambda$ for spectrum q. The weights used here are
\begin{equation}
    \label{eq:lya_weights}
w_{i}=\sigma_{q}^{-2}\left(\lambda_{i}\right)\left(\frac{1+z_{i}}{1+2.25}\right)^{\gamma_{\rm Ly\alpha}-1} ~,
\end{equation}
where $\sigma_{q}^{2}\left(\lambda_{i}\right)$ is the pixel variance due to instrumental noise
and large-scale structure (LSS), and
the redshift evolution of the \lya{} bias is taken into account \citep[$\gamma_{\rm Ly\alpha} = 2.9$,][] {McDonald+2006}.

Correcting the distortion (equation \ref{eq:distortion_correction}) introduces a slight change in the evolution of $\delta\left(\lambda\right)$ per wavelength bin, allowing it to deviate from zero. This property is reintroduced by redefining:
\begin{equation}
    \label{eq:zero_correction}
    \delta_{q}\left(\lambda_{i}\right) \rightarrow \delta_{q}\left(\lambda_{i}\right) - \overline{\delta_{q}\left(\lambda\right)} ~.
\end{equation}

During this procedure, whenever a line of sight contains a DLA, pixels where the DLA reduces the transmission by at least 20\% are masked from the analysis (both the computation of $\delta_q$ and the cross-correlations, see section~\ref{sec:xcf}). For the clustering analysis, we require a DLA sample with high purity, as we described in section~\ref{sec:dla}. However, any missed DLA will impact our measurement of the \lya{} forest. This means that we require a highly complete DLA sample for the masking procedures. Thus, here we mask all the  117,458 DLA candidates, even if they don't meet the purity criteria to make it to our DLA sample. 

We note that SBLAs are not masked in the standard analysis. Not masking these objects implies a high degree of cross-covariance between the SBLA-\lya{} cross-correlation of SBLAs and the \lya{} auto-correlation (see section \ref{sec:joint_fit}). Because of this we also recompute the flux transmission field masking SBLAs. SBLAs are masked at the observed frame by removing all the pixels that meet $\left|\log\left(\lambda\right)-\log\left(\lambda_{\rm abs}\right)\right|\leq2.5\cdot10^{-4}$, where $\lambda_{\rm abs}=\lambda_{\rm Ly\alpha}\left(1+z_{\rm abs}\right)$ is the wavelength of the SBLA. Correlations (and models) computed using the standard \lya{} flux transmission field (i.e. not masking SBLAs) are labelled as {\it standard}. Similarly, correlations (and models) computed using the modified \lya{} flux transmission field with SBLAs masked are labelled as {\it masked}.

\section{Cross-correlations: method and modelling}\label{sec:xcf}
We measure the cross-correlation between the \lya{} forest and the DLAs and the SBLAs. Since the \lya{} transmission field is measured in two spectral regions we have four correlation functions:
\begin{itemize}
    \item \lya{}(\lya{}) $\times$ DLA,
    \item \lya{}(\lyb{}) $\times$ DLA,
    \item \lya{}(\lya{}) $\times$ SBLA,
    \item \lya{}(\lyb{}) $\times$ SBLA,
\end{itemize}
where \lya{}(\lya{}) stands for \lya{} absorption in the \lya{} region. 
As we will see in section~\ref{sec:joint_fit}, in order to perform a joint fit we need to mask SBLAs from the \lya{} forest. Thus, we also recompute the \lya{} autocorrelation and the \lya{}-quasar cross-correlation following the steps in \citep{duMasdesBourboux+2020} but using our masked version of $\delta_q$.

In the following subsections, we describe the measurement of the cross-correlation, the distortion matrix (used when fitting the model), the cross-covariance between samples, and the model we assume for the cross-correlation. The correlation functions and their models are computed following the procedure described in \cite{duMasdesBourboux+2020}. Here, we summarize the procedure to compute the cross-correlation, indicating the particularities of the DLA and SBLA measurements. We refer the reader to \cite{duMasdesBourboux+2020} for more details on this measurement and for the procedure to measure the \lya{} auto-correlation. 

\subsection{Measurement of the cross-correlations}
The correlation functions are defined as
\begin{equation}
    \xi_{A}=\frac{\sum_{(i,j)\in A}w_{i}w_{j}\delta_{i}}{\sum_{(i,j)\in A}w_{i}w_{j}} ~,
\end{equation}
where $i,j$ index a \lya{}-object pixel pair belonging to the $\left(r_{\parallel}, r_{\perp}\right)$ pixel A. Here, the term object stands for a DLA or an SBLA depending on the computed correlation function. The \lya{} weights, $w_{i}$, are computed as stated in equation \ref{eq:lya_weights}. The object weights are defined as
\begin{equation}
    \label{eq:obj_weights}
    w_{j}=\left(\frac{1+z_{j}}{1+2.25}\right)^{\gamma_{\rm obj}-1} ~.
\end{equation}
Here, we take $\gamma_{\rm obj}=0$ as we do not consider any redshift evolution of the biases for the DLAs and the SBLAs. For the DLAs, this choice is motivated by the lack of redshift evolution in the DLA bias reported by \cite{Perez-Rafols+2018a}. 
For the SBLAs, this choice is motivated by the fact that we recover a similar bias (see section~\ref{sec:results}), suggesting they come from the same halo population. When computing the cross-correlation with quasars we use $\gamma_{\rm obj}=1.44$ as in \cite{duMasdesBourboux+2020}.

The cross-correlation is computed for all pixel-object pairs (i, j) of separation parallel and perpendicular to the line of sight $\left(r_{\parallel}, r_{\perp}\right)$ up to 200\hmpc{}, in bins of 4\hmpc{}, excluding pairs corresponding to the same line of sight (the correlation vanishes due to the continuum fitting
procedure). Distances are computed assuming \cite{Planck+2016} cosmology as a fiducial cosmology. 
Note that the cross-correlation is not symmetric with respect to $r_{\parallel}$, and we define it as positive when the object (DLA or SBLA) is in front of the \lya{} pixel, i. e. $z_{\rm Ly\alpha} > z_{\rm obj}$. We compute the correlation function in bins of 4\hmpc, resulting in a total of $N_{\rm bin}=100\times50=5000$ bins.

The covariance matrix is computed using the same estimator as in \cite{duMasdesBourboux+2020}
\begin{equation}
    \label{eq:covariancematrix}
    C_{AB}=\frac{1}{W_{A}W_{B}}\sum_{s}W^{s}_{A}W^{s}_{B}\left(\xi^{s}_{A}\xi^{s}_{B} - \xi_{A}\xi_{B}\right)~,
\end{equation}
where $s$ is a sub-sample with summed weight $W^{s}_{A}$ and measured correlation $\xi^{s}_{A}$, and $W_{A}=\sum_{s}W^{s}_{A}$. Note that here we neglect the small cross-correlation between the subsamples.

\subsection{Cross-covariance between samples}\label{sec:cross_covariance_computation}
We compute the cross-covariance between samples by using a modified version of the estimator of the covariance matrix in equation \ref{eq:covariancematrix}:
\begin{equation}
    C_{AB}^{12}=\frac{1}{W_{A}W_{B}}\sum_{s}W^{s}_{A}W^{s}_{B}\left(\xi^{1,s}_{A}\xi^{2,s}_{B} - \xi^{1}_{A}\xi^{2}_{B}\right)~,
\end{equation}
where, $\xi^{1}$ and $\xi^{2}$ are two different measured correlation functions.

\subsection{Measurement of the distortion matrix}
The projection performed in equations \ref{eq:distortion_correction} and \ref{eq:zero_correction} mix the $\delta_{q}$ within a given forest and modify significantly the correlation function. Assuming that the transformations on $\delta_{q}$ are linear combinations of the true $\delta_{q}$, the true and distorted correlation functions are related by a ``distortion matrix", $D_{AB}$, as
\begin{equation}
    \hat{\xi}_{\rm distorted}\left(A\right) = \sum_{B}D_{AB}\xi_{\rm true}\left(B\right) ~.
\end{equation}

The distortion matrix can be estimated, assuming that the projections in equations \ref{eq:distortion_correction} and \ref{eq:zero_correction} capture the full description, as
\begin{equation}
    D_{AB} = W_{A}^{-1}\sum_{(i, j)\in A} w_{i}w_{j}\sum_{(i', j)\in B} \eta_{ii'} ~.
\end{equation}

\subsection{Model of the correlation function}\label{sec:model}
In this section, we present the theoretical model of the correlation functions used here. In general, we will be using an equivalent model to that of the \lya{}-quasar cross-correlation from \citep{duMasdesBourboux+2020}, but adapted to the particular cases of DLAs/SLBAs. However, in section \ref{sec:joint_fit} we also use the \lya{} autocorrelation and its cross-correlation with quasars, DLAs or SBLAs. For these correlation functions, we use the same model as in \citep{duMasdesBourboux+2020}. We now give a short description of the model for the cross-correlation here, emphasising the differences specific to the DLAs/SBLAs cases, and refer the reader to \citep{duMasdesBourboux+2020} for a more detailed explanation. The model for the \lya{} autocorrelation is similar to that of the cross-correlation and given the importance of cross-correlations to this work, we focus on reviewing the methods. Details of the autocorrelation measurement methodology can be found in section 4  of \citet{duMasdesBourboux+2020}.

The theoretical model for the cross-correlation, $\xi^{t}$ is composed of different correlations:
\begin{equation}
    \xi^{t} = \xi^{{\rm Ly\alpha}\times{\rm obj}} + \sum_{m}\xi^{{\rm obj}\times m}~,
\end{equation}
where $\xi^{{\rm Ly\alpha}\times{\rm obj}}$ is the cross-correlation of our objects (DLA, SBLA or QSO, depending on the measured correlation) with the \lya{} forest and $\xi^{{\rm obj}\times m}$ is the cross-correlation of our objects with other metal absorbers in the \lya{} and \lyb{} spectral regions. 

The components of the correlation function are computed from the Fourier Transform of the tracer bias power-spectrum
\begin{equation}
    \hat{P}(\mathbf{k}) = b_{i}b_{j}\left(1+\beta_{i}\mu_{k}^{2}\right) \left(1+\beta_{j}\mu_{k}^{2}\right)P_{\rm QL}(\mathbf{k}) F_{\rm NL}(\mathbf{k})G(\mathbf{k})~,
\end{equation}
where the vector $\mathbf{k}=\left(k_{\parallel}, k_{\perp}\right)=\left(k, \mu_{k}\right)$ has components $\left(k_{\parallel}, k_{\perp}\right)$ parallel and perpendicular to the line-of-sight, with $\mu_{k}0k_{\parallel}/k$. The first multiplicative terms after the equality stand for the bias ($b$) and redshift-space distortion ($\beta$) parameters for tracers $i$ and $j$. $P_{\rm QL}$ is the quasi-linear power spectrum decoupling the peak and smooth component and including non-linear broadening for the BAO peak as explained in \cite{duMasdesBourboux+2020}. $F_{\rm NL}$ corrects for non-linear effects at large $\mathbf{k}$, and $G\left(\mathbf{k}\right)$ corrects for the averaging of the correlation function in individual $\left(r_{\parallel}, r_{\perp}\right)$ bins.

Finally, the correlation function (and the quasi-linear power spectrum accordingly) is decomposed into smooth and peak components, $\xi_{\rm sm}$ and $\xi_{\rm peak}$ respectively, to allow the position of the BAO peak to vary: 
\begin{equation}
    \xi\left(r_{\parallel}, r_{\perp}, \alpha_{\parallel}, \alpha_{\perp}\right) = \xi_{\rm sm}\left(r_{\parallel}, r_{\perp}\right)+ \xi_{\rm peak}\left(\alpha_{\parallel}r_{\parallel}, \alpha_{\perp}r_{\perp} \right) ~.
\end{equation}
The variation is characterized by the BAO parameters $\alpha_{\parallel}$ and $\alpha_{\perp}$:
\begin{equation}
    \alpha_{\parallel} = \frac{\left[D_{H}\left(z_{\rm eff}\right)/r_{d}\right]}{\left[D_{H}\left(z_{\rm eff}\right)/r_{d}\right]_{\rm fid}}\,\,\,\text{and}\,\,\,\alpha_{\perp} = \frac{\left[D_{M}\left(z_{\rm eff}\right)/r_{d}\right]}{\left[D_{M}\left(z_{\rm eff}\right)/r_{d}\right]_{\rm fid}} ~,
\end{equation}
where $D_{H}$ and $D_{M}$ are the Hubble and coming angular diameter distances at the redshift of references $z_{\rm eff}$, and $r_{d}$ is the scale of the sound horizon.

To account for unknown astrophysical effects and other possible sources of systematic errors, we add polynomial "broadband" terms to the cross-correlations between the \lya{} forest and SBLAs. We follow the same choice as \cite{duMasdesBourboux+2020, Bautista+2017} and adopt the form 
\begin{equation}
    B(r, \mu) = \sum_{j=0}^{j_{\rm max}}\sum_{i=i_{\rm min}}^{i_{\rm max}}a_{ij}\frac{L_{j}\left(\mu\right)}{r^{i}}\text{ ($j$ even)}~,
\end{equation}
where $L_{j}$ are the Legendre polynomials and we take $\left(i_{\rm min}, i_{\rm max}\right)=(0, 2)$, and $j_{\rm max}=6$. The broadbands are only added to the correlations involving SBLAs as correlations involving DLAs are too noisy for broadbands to be relevant. The correlations involving quasars and the \lya{} auto-correlation are shown to not need these broadbands in \cite{duMasdesBourboux+2020}.

The standard \lya{} BAO fits include data points from 10 to 180\hmpc{}. The reason for the lower cut is that the linear approximation starts to break at small distances, where both non-linearities and astrophysical processes start to become more and more relevant. The larger cut is set to keep the computing time manageable. Since the BAO peak occurs at roughly $\sim100$\hmpc{}, this large distance cut is sufficiently large for our measurements. Here we deviate from these standard \lya{} BAO fits by discarding object-pixel pair data where galaxies in absorption are close to the line-of-sight such that  $r_{\perp} > 30$\hmpc{}. We discuss the reasons behind this requirement in section~\ref{sec:joint_fit}.
\section{Cross-correlation  Measurements}\label{sec:results}
In this section, we present the measured cross-correlation functions for DLAs and SBLAs and their best-fit models. For clarity, throughout this section we show only the cross-correlations with the \lya{} absorption in the \lya{} region, i.e. \lya{}(\lya{}) $\times$ DLA
and  \lya{}(\lya{}) $\times$ SBLA. However, the reported results for the fits are computed in a joint fit using also the \lya{}(\lyb{}) cross-correlations. The \lya{}(\lyb{}) correlations are given in appendix~\ref{sec:lya(lyb)}. Generally, we will refer to the DLA and SBLA bias and redshift space distortions as $b_{\rm gal\, abs}$ and $\beta_{\rm gal\, abs}$ to reflect that both objects are identified through their absorption of background light and are expected to be associated with galaxies.

\subsection{\lya{}-DLA cross-correlation}\label{sec:dla}
\begin{figure}
    \centering
    \includegraphics[width=\columnwidth]{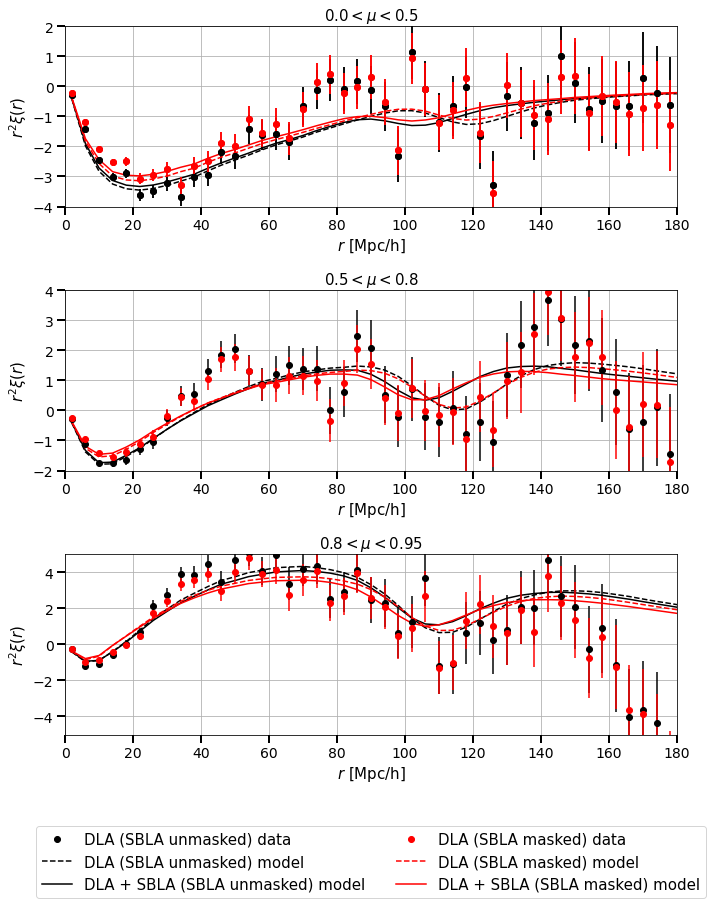}
    \caption{\lya{}(\lya{}) $\times$ DLA correlation function (points with errorbars). The best fit model is overplotted as a dashed line and was computed using a joint fit including also \lya{}(\lyb{}) $\times$ DLA correlation function. Black points and lines show the data and best-fit model using the standard \lya{} flux transmission field, i.e. not masking SBLAs. Red points and lines show the data and best-fit model with masked \lya{} flux transmission field. Solid lines show the best-fit model including the cross-correlation of \lya{} with both DLAs and SBLAs (on data with masked flux transmission field). The different panels show different bins of $\mu=r_{\parallel}/r$ from perpendicular to the line-of-sight (top) to close to the line-of-sight (bottom). We can see that masking SBLAs does not significantly impact the \lya{}(\lya{}) $\times$ DLA cross-correlation.}
    \label{fig:dlaxlya}
\end{figure}

Figure~\ref{fig:dlaxlya} shows the \lya{}(\lya{}) $\times$ DLA cross-correlation for both a masked and unmasked flux transmission field
The best-fit model to each are shown as solid lines and its parameters are given in the 
table~\ref{tab:best_fit_param}. The recovered value for the DLA bias is $2.336\pm0.058$ when SBLAs are not masked in the computation of the \lya{} flux transmission field and $2.337\pm0.058$ when they are. It is notable that there is no effect on the DLA bias when we mask or do not mask the SBLAs from the computation of the \lya{} flux transmission field.

\cite{Perez-Rafols+2018a} find a slightly lower value of $2.00\pm0.19$ for the DLA bias. This shift in the DLA bias might be explained by the different DLA samples. The sample used in \cite{Perez-Rafols+2018a} was the DR12 extension of the DLA catalogue from \cite{Noterdaeme+2012}, which contains DLA candidates and is more focused on completeness. If we assume that most DLA contaminants are typical Lyman alpha forest pixels, then we expect the mean bias to be lower in a sample which is less pure. Thus, the higher value seems suggestive that the purity of our sample is higher. However, the reality is probably more complex and some contaminants could have a higher bias. This means that the effective bias we measure here is a complex function that depends on the bias of the different contaminants. A more detailed study comparing these samples would be desired, but we leave this for the future. We note, however, that changes in the bias do not affect our ability to measure the BAO parameters, the main focus of this work. 


\begin{table*}
    \centering
\scalebox{0.9}{
\begin{tabular}{l|lll|lll}
\toprule
    & \multicolumn{3}{c|}{SBLAs unmasked} & \multicolumn{3}{c}{SBLAs masked} \\
    Parameters &      DLA &     SBLA & DLA + SBLA &        DLA &       SBLA & DLA + SBLA  \\
    \midrule
    $\alpha_{\parallel}$ &  $0.950\pm0.078$ &      $0.963\pm0.085$ &          $0.951\pm0.059$ & $0.950\pm0.075$ &    $1.030\pm0.087$ &        $0.977\pm0.067$ \\
    $\alpha_{\perp}$ &  $0.848\pm0.058$ &      $1.033\pm0.104$ &          $0.984\pm0.069$ & $0.850\pm0.070$ &    $1.001\pm0.090$ &        $0.994\pm0.070$ \\
    $b_{\eta,{\rm Ly\alpha}}$ &   $-0.161\pm0.024$ &     $-0.190\pm0.013$ &         $-0.193\pm0.009$ &   $-0.183\pm0.024$ &   $-0.182\pm0.011$ &       $-0.183\pm0.010$ \\
    $\beta_{\rm Ly\alpha}$ &      $2.04\pm0.10$ &        $2.06\pm0.09$ &            $2.07\pm0.09$ &      $2.04\pm0.10$ &      $2.05\pm0.09$ &          $2.05\pm0.09$ \\
    $b_{\eta,{\rm gal\,abs}}$ &              $1.0$ &                $1.0$ &                    $1.0$ &              $1.0$ &              $1.0$ &                  $1.0$ \\
    $\beta_{\rm gal\,abs}$ &    $0.415\pm0.010$ &      $0.417\pm0.010$ &          $0.418\pm0.010$ &    $0.415\pm0.010$ &    $0.416\pm0.010$ &        $0.416\pm0.010$ \\
    $\Delta r_{\parallel,{\rm gal\,abs}} \left({\rm h^{-1}Mpc}\right)$ &       $-2.0\pm1.8$ &          $0.7\pm0.7$ &              $0.4\pm0.6$ &       $-2.0\pm1.7$ &        $0.6\pm0.7$ &            $0.3\pm0.7$ \\
    $\sigma_{v,{\rm gal\,abs}} \left({\rm h^{-1}Mpc}\right)$ &       $0.0\pm14.9$ &         $0.0\pm14.7$ &             $0.0\pm10.9$ &       $0.1\pm14.7$ &       $0.0\pm13.0$ &            $0.0\pm7.6$ \\
    $10^{3}b_{\eta,{\rm SiII(126)}}$ &          $-0\pm18$ &            $-15\pm5$ &                $-15\pm4$ &          $-0\pm20$ &          $-14\pm5$ &              $-11\pm5$ \\
    $10^{3}b_{\eta,{\rm SiIII(120.7)}}$ &           $-0\pm4$ &             $-0\pm2$ &                 $-0\pm1$ &           $-0\pm5$ &           $-0\pm4$ &               $-0\pm2$ \\
    $10^{3}b_{\eta,{\rm SiII(119.3)}}$ &          $-0\pm13$ &            $-0\pm13$ &                 $-0\pm2$ &          $-0\pm13$ &           $-0\pm3$ &               $-0\pm2$ \\
    $10^{3}b_{\eta,{\rm SiII(119)}}$ &          $-0\pm13$ &            $-2\pm10$ &                 $-0\pm4$ &          $-0\pm14$ &          $-0\pm12$ &               $-0\pm3$ \\
    $10^{3}b_{\eta,{\rm CIV({\rm eff})}}$ & $-0.0050\pm0.0026$ &   $-0.0050\pm0.0025$ &       $-0.0050\pm0.0025$ & $-0.0050\pm0.0026$ & $-0.0050\pm0.0025$ &     $-0.0050\pm0.0025$ \\
    $b_{\rm HCD}$ &           $-0.035$ &             $-0.035$ &                 $-0.035$ &           $-0.000$ &           $-0.000$ &               $-0.000$ \\
    $\beta_{\rm HCD}$ &            $0.493$ &              $0.493$ &                  $0.493$ &            $0.500$ &            $0.500$ &                $0.500$ \\
    \midrule
    $\chi^{2}$ &           5021.390 &             5162.961 &                10190.594 &           5019.467 &           5141.161 &              10164.458 \\
    $N_{\rm bin}$ &               4940 &                 4940 &                     9880 &               4940 &               4940 &                   9880 \\
    $N_{\rm param}$ &                 12 &                   36 &                       36 &                 12 &                 36 &                     36 \\
    probability &              0.173 &                0.005 &                    0.007 &              0.178 &              0.009 &                  0.012 \\
    \midrule
    $b_{\rm Ly\alpha}$ &  $-0.077\pm-0.004$ &    $-0.090\pm-0.004$ &        $-0.091\pm-0.004$ &  $-0.087\pm-0.004$ &  $-0.086\pm-0.004$ &      $-0.086\pm-0.004$ \\
    $b_{\rm gal\,abs}$ &    $2.336\pm0.058$ &      $2.329\pm0.057$ &          $2.321\pm0.056$ &    $2.337\pm0.058$ &    $2.334\pm0.057$ &        $2.333\pm0.057$ \\
\bottomrule
\end{tabular}
}
    \caption{Best-fit parameters for the correlation functions using DLAs and SBLAs. Values without error bars were fixed to the given values. Error bars are given by \texttt{iminuit} for all parameters except for $\alpha_{\parallel}$ and $\alpha_{\perp}$. These two parameter errors are estimated from a $\chi^2$ scan. The first block contains measurements using the standard \lya{} flux transmission field for DLA and SBLA. The second block contains the same measurements using masked \lya{} flux transmission field plus a joint fit of the two. Here DLAs and SBLAs are assumed to trace the density field in a similar way as galaxies in absorption with common bias and redshift space distortions. Their specific parameter is labelled as ${\rm gal\, abs}$. In order to more easily compare with results from \protect\cite{Perez-Rafols+2018a}, the \lya{} forest bias and the bias of these systems are derived at the bottom of the table (but not used in the fit). Note that each of the fits uses the correlations from the \lya{} and \lyb{} regions.}
    \label{tab:best_fit_param}
\end{table*}

\subsection{\lya{}-SBLA cross-correlation}\label{sec:sbla}
\begin{figure}
    \centering
    \includegraphics[width=\columnwidth]{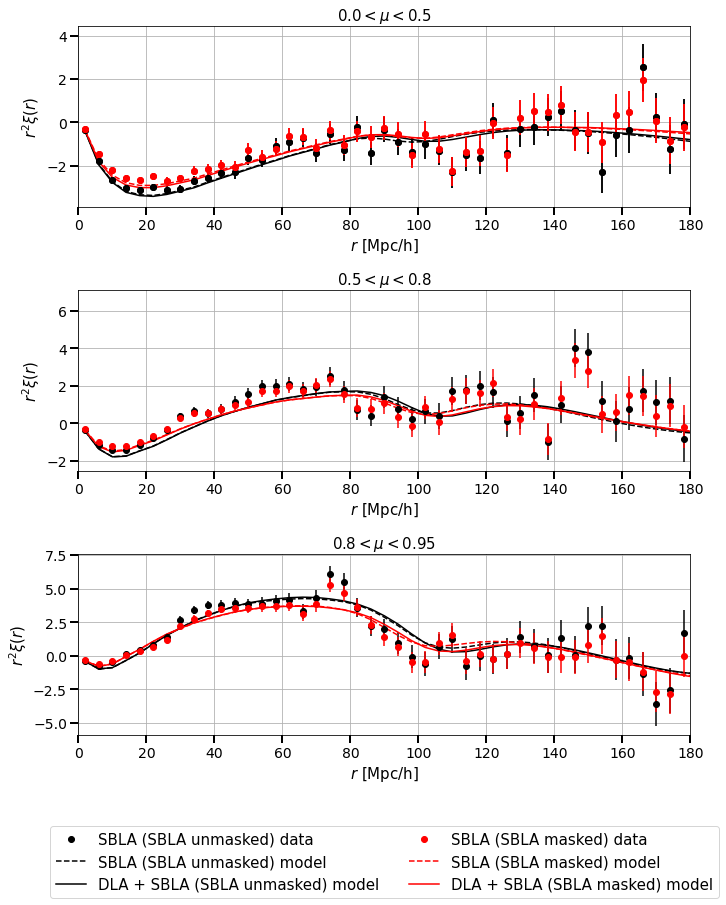}
    \caption{Same as figure~\ref{fig:dlaxlya} but for the  \lya{}(\lya{}) $\times$ SBLA correlation function. We can see that masking SBLAs has a larger impact here (compared to \lya{}(\lya{}) $\times$ DLA). The correlation function is affected mostly in distances $<$80\hmpc{}.}
    \label{fig:sblaxlya}
\end{figure}

Having analysed the correlation between DLAs and the \lya{} forest we now shift our attention to SBLAs. The black points in Figure~\ref{fig:sblaxlya} show the \lya{}(\lya{}) $\times$ SBLA cross-correlation using the standard flux transmission field. The best-fit model is shown as black solid lines and its parameters are given in the first block of  table~\ref{tab:best_fit_param}. In the plot, we can see that the errors of the correlation function are much smaller compared to DLAs, as expected from the fact that the SBLA sample is much larger. The measured average SBLA correlation function errors are 32\% of the DLA correlation function errors, corresponding to finding 15 times more pairs of SBLA-\lya{} pixels than DLA-\lya{} pixels. We note that this is different from the ratio of objects in the samples as it depends in a non-trivial way on the geometry of the survey.

The measured SBLA bias is $2.329\pm0.057$ when SBLAs are not masked in the computation of the \lya{} flux transmission field and $2.334\pm0.057$ when they are. This value is consistent with the measured DLA bias, supporting our assumption that both systems originate in similar physical systems. 

The similarity of the bias parameters for SBLAs and DLAs suggests that both systems reside in a similar halo population. There is also no effect on the SBLA bias when we mask or do not mask the SBLAs from the computation of the \lya{} flux transmission field. 

\subsection{Combining \lya{}-DLA and \lya{}-SBLA cross-correlations}
We have found the bias and $\beta$ parameters for DLAs and SBLAs to be similar, suggesting suggesting that they reside in a similar halo population. If this is correct, then we would expect the two correlation functions to share the same parameters. To confirm this, we perform a joint fit of the DLAs and SBLAs cross-correlations. The fit results are shown in table~\ref{tab:best_fit_param}, and the best-fit models are plotted in figures~\ref{fig:dlaxlya} and \ref{fig:sblaxlya}. We can see that the recovered values are consistent with the fits to DLAs and SBLAs separately. We see that the joint fit is dominated by the SBLA cross-correlations as expected by its larger signal-to-noise. However, the joint fit does help in tightening the constraints on the different parameters. The recovered values of $b_{\rm gal\, abs}$ and $\beta_{\rm gal\, abs}$ are also consistent with the values found in the individual fits. 
\section{The value of additional cross-correlations}\label{sec:joint_fit}
In Section~\ref{sec:results} we have shown the first clustering measurements from galaxies detected in absorption (namely DLAs and SBLAs) by correlating them with the \lya{} forest. Naturally, the question arises: how useful are these cross-correlations compared to the well-established \lya{} auto-correlation and the \lya{} $\times$ QSO cross-correlation? In this section, we explore the answer to this question and justify the need for realistic mocks if these new correlations are to be added to the standard \lya{} BAO analysis.

We start by studying the limitations of a joint fit in section~\ref{sec:cross_covariance}, which is part of our motivation to explore masking SBLAs in the \lya{} forest. We then explore the effect this masking has on the different correlation functions in section~\ref{sec:masking_effects}. Then, we assess the relative importance of the new additions by performing a joint fit in section~\ref{sec:rel_importance}. We warn the reader that the \lya{} $\times$ GalAbs cross-correlation is not tested with mocks so it may be potentially subject to systematics. Thus, the BAO errors and results should be taken with a grain of salt and only as a proof of concept of what could be achieved. In this work we focus on the estimates of BAO uncertainty that we can gain from the data alone, and deliberately neglect the BAO constraints themselves.

Finally, it is worth noting that the fits performed here ignore the bins of the correlation functions with $r_{\perp, {\rm min}} < 30$\hmpc{} (see section~\ref{sec:model}). We discuss this choice in section~\ref{sec:rtmin}. Note that to have a clean apples-to-apples comparison when SBLAs are masked in the flux transmission field, this cut is applied to all the correlation functions. This means that the results for the \lya{} auto-correlation and the \lya{} $\times$ QSO cross-correlation in tables~\ref{tab:lya_fits} and \ref{tab:joint_fit_param} will not be exactly the same as those reported by \cite{duMasdesBourboux+2020}.  
 
\subsection{Cross-covariance between the samples}\label{sec:cross_covariance}
The standard \lya{} BAO analysis \citep{duMasdesBourboux+2020} performs a joint fit using four correlation functions \lya{}(\lya{}) $\times$ \lya{}(\lya{}),  \lya{}(\lya{}) $\times$ \lya{}(\lyb{}), \lya{}(\lya{}) $\times$ QSO and \lya{}(\lyb{}) $\times$ QSO. In this joint fit, the cross-covariance between these correlation functions is neglected. This is necessary as the covariance matrices for the individual samples are computed via the sub-sampling technique, and there are not enough sub-samples to compute the full covariance matrix including the cross-covariance between the samples. A direct calculation of the full covariance matrix including the cross-covariance between the samples is currently too computationally expensive. By neglecting the cross-covariance, the covariance matrices for the individual correlation functions can be inverted independently and a joint fit can be performed in a reasonable computing time. 

\begin{figure*}
    \centering
    \includegraphics[width=0.45\textwidth]{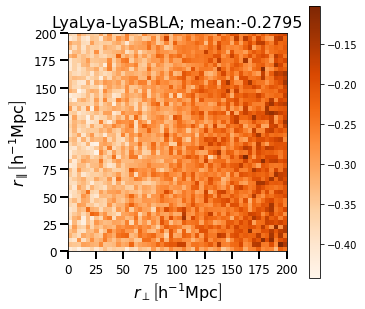}
    \hspace{0.05\textwidth}
    \includegraphics[width=0.45\textwidth]{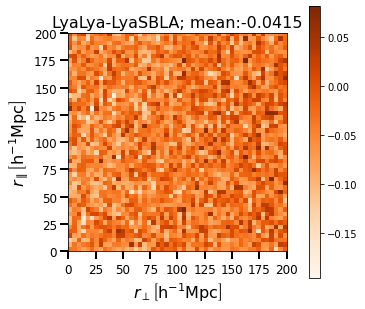}
    \\
    \includegraphics[width=0.45\textwidth]{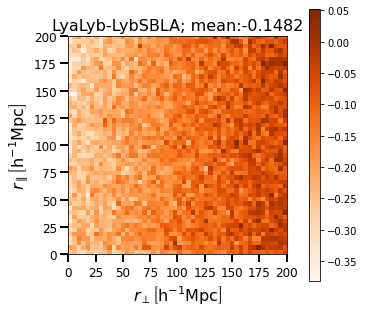}
    \hspace{0.05\textwidth}
    \includegraphics[width=0.45\textwidth]{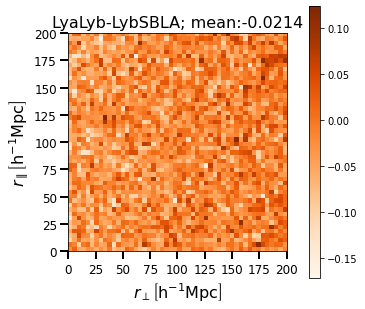}
    \caption{Top: cross-covariance between the correlations \lya{}(\lya{}) $\times$ \lya{}(\lya{}) and \lya{}(\lya{}) $\times$ SBLA when SBLAs are not masked (left) or are masked (right) in the computation of the \lya{} flux transmission field. Bottom panels: cross-covariance between the correlations \lya{}(\lya{}) $\times$ \lya{}(\lyb{}) and \lya{}(\lyb{}) $\times$ SBLA when SBLAs are not masked (left) or are masked (right) in the computation of the \lya{} flux transmission field}
    \label{fig:cross_covariance}
\end{figure*}

\cite{duMasdesBourboux+2020} showed that the cross-covariance between the \lya{} auto-correlation and its cross-correlation with quasars is very small and can thus be neglected (see their figure 11). Here, we compute the cross-covariance between pairs of correlations (see section~\ref{sec:cross_covariance_computation}) to check whether or not they are also small. We find that this cross-covariance is generally negligible except for two cases: i) the correlations \lya{}(\lya{}) $\times$ \lya{}(\lya{}) and \lya{}(\lya{}) $\times$ SBLA (see top left panel in figure~\ref{fig:cross_covariance}), where we see a clear cross-covariance with a mean of $\sim-30\%$, and ii) the correlations \lya{}(\lya{}) $\times$ \lya{}(\lyb{}) and \lya{}(\lyb{}) $\times$ SBLA (see bottom left panel in figure~\ref{fig:cross_covariance}), where we see a clear cross-covariance with a mean of $\sim-15\%$. This is expected as SBLAs are both present in the \lya{} flux transmission field and used as tracers. The negative sign is also expected as they enter the sample with a negative bias when part of the \lya{} flux transmission field and with a positive bias when acting as tracers. 

These cross-covariances are greatly reduced to $<5\%$ by masking these SBLAs from the \lya{} forest (see right panels in figure~\ref{fig:cross_covariance}). The values of the cross-covariance when SBLAs are masked are small enough that a joint fit is possible (though only marginally). As the statistical constraints become tighter and tighter, this approximation might no longer be valid and so new fitting strategies need to be explored to be prepared for this future challenge. An improved DLA and SBLA identification algorithm would also help mitigate this problem. The validation of such an algorithm would require the usage of simulated spectra including these objects, that are unfortunately not available to date.

\subsection{Effects of masking SBLAs in the \lya{} forest}\label{sec:masking_effects}
We have seen that in order to make the cross-covariance negligible to perform a joint analysis we need to mask the SBLAs from the \lya{} flux transmission field. Here we assess the impact this has on the different correlation functions and its constraining power on the BAO scales. In section~\ref{sec:results}, we have already discussed the effect of masking these objects in the \lya{} $\times$ DLA and \lya{} $\times$ SBLA correlation functions. Here we discuss the impact this masking has on the \lya{} auto-correlation and its cross-correlation with quasars.

We expect the constraining power of these correlation functions to be slightly worse since we have removed some of the largest absorbers from the forest. Table~\ref{tab:lya_fits} show the best-fit results for the \lya{} auto-correlation, the \lya{} $\times$ QSO cross-correlation and the joint fit of both with and without masking SBLAs. Surprisingly, we find that the BAO constraints are essentially unaffected. 

We do observe, however, a decrease of 0.012 ($2.2\sigma$) in the recovered \lya{} forest bias (in absolute value) when masking SBLAs and using only the \lya{} forest autocorrelation. When the cross-correlation with quasars is also included, the decrease is 0.012 ($2.4\sigma$). The change goes in the expected direction since we removed the points with larger bias from the analysis. 

When considering only the cross-correlation with quasars, we observe an increase in the \lya{} bias. However, we observe that when we are masking SBLAs the recovered value for the \lya{} bias is more in agreement with the measurements from the \lya{} auto-correlation alone and when both correlations are used in the fit. We also observe that there is a significant decrease in the quasar bias error (the error when masking SBLAs is $\sim13$ times smaller). This suggests that it is also important to mask these objects when measuring the cross-correlation with quasars.

\begin{table*}
    \centering
\scalebox{0.9}{
\begin{tabular}{l|lll|lll}
\toprule
    & \multicolumn{3}{c|}{SBLAs unmasked} & \multicolumn{3}{c}{SBLAs masked} \\          Parameters &  \lya{} &     QSO & \lya{}+QSO  &             \lya{} &                QSO &         \lya{}+QSO \\
    \midrule
    $\alpha_{\parallel}$ &  $1.057\pm0.041$ &  $1.035\pm0.040$ &      $1.043\pm0.029$ & $1.050\pm0.038$ & $1.044\pm0.042$ &    $1.044\pm0.027$ \\
    $\alpha_{\perp}$ &  $0.976\pm0.048$ &  $0.955\pm0.040$ &      $0.964\pm0.030$ & $0.974\pm0.046$ & $0.946\pm0.043$ &    $0.959\pm0.030$ \\
    $b_{\eta,{\rm Ly\alpha}}$ &   $-0.236\pm0.005$ &   $-0.172\pm0.089$ &     $-0.217\pm0.011$ &   $-0.210\pm0.005$ &   $-0.210\pm0.007$ &   $-0.193\pm0.010$ \\
    $\beta_{\rm Ly\alpha}$ &      $2.02\pm0.09$ &      $2.04\pm0.10$ &        $2.02\pm0.10$ &      $2.03\pm0.09$ &      $2.03\pm0.10$ &      $2.03\pm0.09$ \\
    $b_{\eta,{\rm QSO}}$ &                    &             $1.00$ &               $1.00$ &                    &             $1.00$ &             $1.00$ \\
    $\beta_{\rm QSO}$ &                    &    $0.199\pm0.105$ &      $0.255\pm0.011$ &                    &    $0.244\pm0.012$ &    $0.261\pm0.016$ \\
    $\Delta r_{\parallel,{\rm QSO}} \left({\rm h^{-1}Mpc}\right)$ &                    &   $-0.013\pm0.585$ &     $-0.049\pm0.585$ &                    &   $-0.240\pm0.621$ &   $-0.291\pm0.628$ \\
    $\sigma_{v,{\rm QSO}} \left({\rm h^{-1}Mpc}\right)$ &                    &    $5.236\pm3.715$ &      $6.155\pm1.871$ &                    &    $8.941\pm2.413$ &    $6.565\pm2.470$ \\
    $10^{3}b_{\eta,{\rm SiII(126)}}$ &           $-6\pm4$ &           $-4\pm4$ &             $-4\pm3$ &           $-4\pm3$ &           $-4\pm4$ &           $-3\pm2$ \\
    $10^{3}b_{\eta,{\rm SiIII(120.7)}}$ &           $-7\pm3$ &           $-0\pm2$ &            $-0\pm14$ &           $-7\pm3$ &           $-0\pm3$ &          $-0\pm13$ \\
    $10^{3}b_{\eta,{\rm SiII(119.3)}}$ &           $-5\pm4$ &           $-0\pm2$ &             $-1\pm2$ &           $-6\pm3$ &           $-0\pm3$ &           $-2\pm3$ \\
    $10^{3}b_{\eta,{\rm SiII(119)}}$ &          $-0\pm16$ &           $-0\pm2$ &             $-0\pm4$ &          $-0\pm15$ &           $-0\pm3$ &           $-0\pm3$ \\
    $10^{3}b_{\eta,{\rm CIV({\rm eff})}}$ & $-0.0051\pm0.0026$ & $-0.0050\pm0.0026$ &   $-0.0051\pm0.0026$ & $-0.0051\pm0.0026$ & $-0.0050\pm0.0026$ & $-0.0051\pm0.0026$ \\
    $b_{\rm HCD}$ &   $-0.000\pm0.146$ &           $-0.035$ &     $-0.035\pm0.018$ &   $-0.000\pm0.145$ &           $-0.000$ &   $-0.033\pm0.016$ \\
    $\beta_{\rm HCD}$ &    $0.500\pm0.090$ &            $0.493$ &      $0.493\pm0.088$ &    $0.500\pm0.090$ &            $0.500$ &    $0.499\pm0.090$ \\
    $10^{2}A_{\rm sky, Ly\alpha(Ly\beta)}$ &    $0.009\pm0.001$ &                    &      $0.009\pm0.001$ &    $0.006\pm0.001$ &                    &    $0.006\pm0.001$ \\
    $\sigma_{\rm sky, Ly\alpha(Ly\beta)}$ &       $30.9\pm2.4$ &                    &         $30.9\pm1.9$ &       $33.8\pm3.2$ &                    &       $33.9\pm3.2$ \\
    \midrule
    $\chi^{2}$ &           2492.818 &           5029.771 &             7525.161 &           2440.553 &           5005.963 &           7449.698 \\
    $N_{\rm bin}$ &               2470 &               4940 &                 7410 &               2470 &               4940 &               7410 \\
    $N_{\rm param}$ &                 15 &                 12 &                   18 &                 15 &                 12 &                 18 \\
    probability &              0.292 &              0.153 &                0.137 &              0.578 &              0.215 &              0.316 \\
    \midrule
     $b_{\rm Ly\alpha}$ &  $-0.113\pm-0.005$ &  $-0.082\pm-0.004$ &    $-0.104\pm-0.005$ &  $-0.101\pm-0.005$ &  $-0.100\pm-0.005$ &  $-0.092\pm-0.004$ \\
     $b_{\rm QSO}$ &                    &    $4.884\pm2.569$ &      $3.799\pm0.167$ &                    &    $3.978\pm0.196$ &    $3.718\pm0.229$ \\
\bottomrule
\end{tabular}
}
    \caption{Best-fit parameters for the \lya{} auto-correlation (\lya{}), its cross-correlation with quasars (QSO) and a joint fit of the two (\lya{}+QSO). The first block contains the fits when SBLAs are not masked from the \lya{} flux transmission field and the second block when they are. Values without error bars were fixed to the given values and empty values indicate the parameters that were not used in some of the fits. Error bars are given by \texttt{iminuit} for all parameters except for $\alpha_{\parallel}$ and $\alpha_{\perp}$. These two parameter errors are estimated from a $\chi^2$ scan. The \lya{} forest and quasar biases are derived at the bottom of the table (but not used in the fit). We see that masking SBLAs from the computation of the \lya{} flux transmission field does not hurt the constraining power of the correlation functions. Note that the results on this table do not match those of \protect\cite{duMasdesBourboux+2020} as our fits do not include the bins with $r_{\perp} < 30$\hmpc{} (see text for details).}
    \label{tab:lya_fits}
\end{table*}

\subsection{Choice of the \texorpdfstring{$r_{\perp}$}{rt} cut}\label{sec:rtmin}
\begin{figure}
    \centering
    \includegraphics[width=\columnwidth]{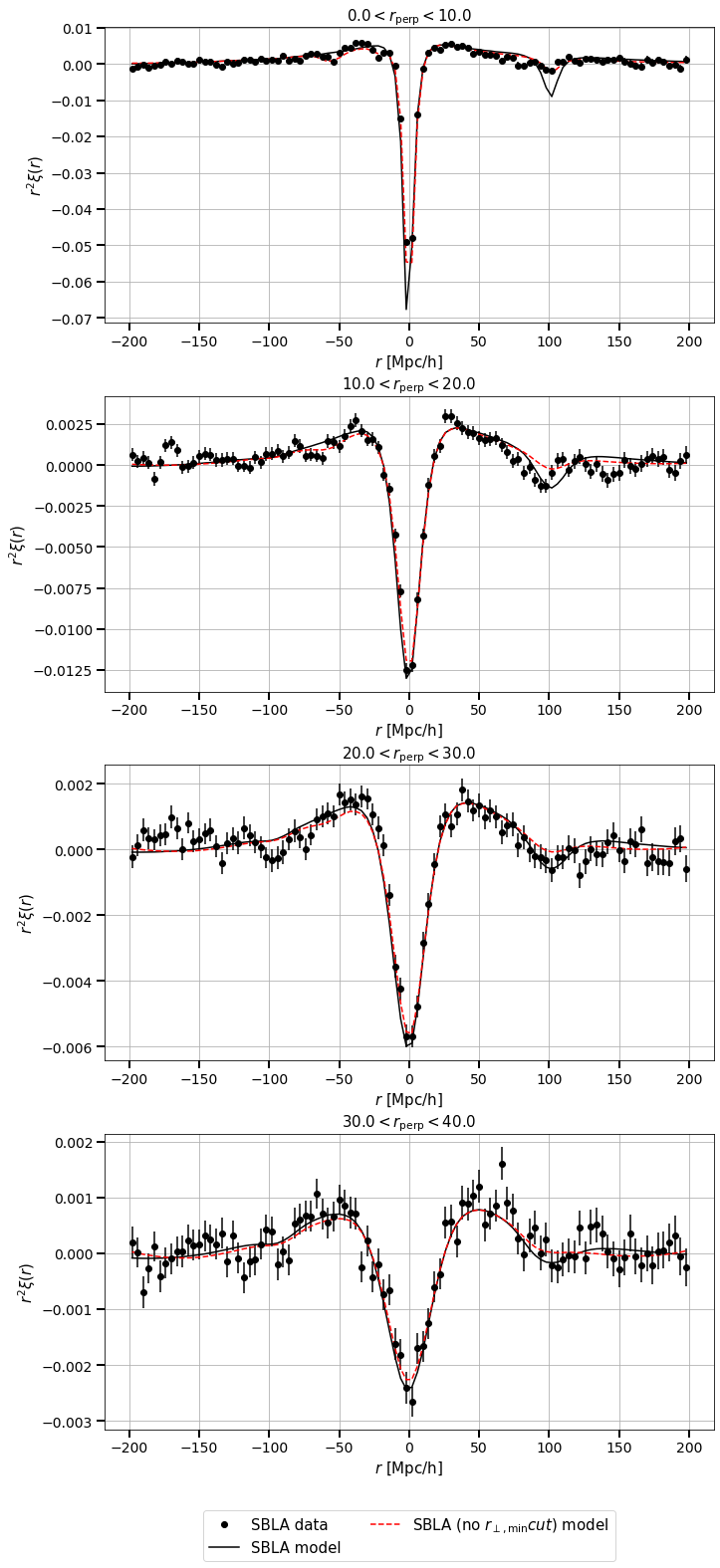}
    \caption{\lya{}(\lya{}) $\times$ SBLA correlation function vs $r_{\parallel}$ in bins of $r_{\perp}$ (values are given in \hmpc{}) when SBLAs are masked in the computation of the \lya{} flux transmission field (points with errorbars). The black solid line corresponds to our best-fit model and the red dashed line corresponds to the best-fit model where the bins with $r_{\perp}<30$ are also included in the fit. We see that the models underpredict the Si\thinspace{}II(119.0) and overpredict the Si\thinspace{}III(126.0) by a factor of 3.}
    \label{fig:sblaxlya2}
\end{figure}

In section~\ref{sec:model} we explained that we exclude those bins in the correlation function that have $r_{\perp}<30$\hmpc{}. Here we justify this choice. We re-compute the fits to the \lya{} $\times$ SBLA cross-correlation using the \lya{} flux transmission field with masked SBLA and not including this extra constraint.  The results of this exercise can be seen in figure~\ref{fig:sblaxlya2}, where we plot the \lya{}(\lya{}) $\times$ SBLA cross-correlation, our selected model with $r_{\perp}<30$\hmpc{}, and the model without this cut. The different panels show the correlation function as a function of $r_{\parallel}$ for increasing bins in $r_{\perp}$.
In the first $r_{\perp}$ bin, we see that the fit model underpredicts the Si\thinspace{}II(119.0) and overpredicts the Si\thinspace{}III(126.0) by a factor of 3. The problem alleviates as we separate further from the quasar and is no longer seen in the bin for $r_{perp}>30$\hmpc{}.


Let us focus on the bin closest to the line of sight. This is the bin that is most affected by the contamination by metals. While we model our contamination, this plot suggests that our model for the metal contamination is not good enough to explain the observations. We note some key differences between the SBLAs and the other tracers (\lya{} forest, DLAs, and QSOs) that could explain why correlations involving SBLAs are more sensitive to metal contamination. 

First, we expect most of the metals to be closer to galaxies as they are mostly generated by supernovae from stars on those galaxies. Because we are now using these as tracers, it is possible that we are more sensitive to non-linearities present in the metal enrichment of the gas. We would not see this effect on the correlations with DLAs as our statistical precision is not good enough. On top of this, the redshift accuracy of SBLAs is far superior to that of quasars and DLAs. Quasar redshifts are identified using the emission lines, which are known to have systemic shifts \citep[see e.g.][]{Shen+2016}. Instead, the redshift of SBLAs is directly determined by the absorption of the \lya{} line. This can also be seen comparing the parameters $\sigma_{v,{\rm QSO}}$ and $\sigma_{v,{\rm gal}}$ from tables~\ref{tab:best_fit_param} and \ref{tab:lya_fits}. The former yields $\sim10$\hmpc{} whereas the latter yields $\sim0$\hmpc{}.

Other possible explanations for this effect include unknown astrophysical systematic effects that bleed to the 3D correlation function, target selection effects and/or other sources of systematic errors. By not including bins with $r_{\perp}<30$\hmpc{} in the fit we are mitigating these effects. To reliably include these low perpendicular separations in our parameter fits (and thus improve our constraining power) we would require a more in-depth study of this metal modelling including tests with synthetic spectra. We leave such a study for the future when said mocks are available.

\subsection{Performance of DLAs/SBLAs compared to other \lya{} BAO probes}\label{sec:rel_importance}
To assess the relative importance of the correlations \lya{} $\times$ GalAbs compared to \lya{} auto and \lya{} $\times$ QSO we compare the BAO constraints of the different probes using the masked flux transmission field. We start by comparing the single tracer fits. Here, we consider DLAs and SBLAs as a single tracer: galaxies in absorption (GalAbs). Even though there are more SBLAs than DLAs, we see in their joined fit in table~\ref{tab:best_fit_param} that the BAO constraints benefit from using both. We also checked that this statement holds even when we consider joined fits with multiple tracers.

The BAO constraints along the line-of-sight from the \lya{} $\times$ GalAbs ($\alpha_\parallel$) are $\sim75\%$ worse than the \lya{} auto-correlation and $\sim60\%$ worse than the \lya{} $\times$ QSO cross-correlation. For the transverse BAO parameter ($\alpha_\perp$), the constraints are $\sim50\%$ worse than for the \lya{} auto-correlation and $\sim60\%$ worse than for the \lya{} $\times$ QSO cross-correlation. We see that this measurement is not competitive on its own. However, we stress that the correlation with galaxies is more affected by an imperfect metal contamination model which motivated the exclusion of low $r_{\perp}$ bin from the fit (see section~\ref{sec:rtmin}). The relative importance of this tracer should be reassessed once a better metal contamination model is developed.

We now move to explore the impact on joint fits. We perform joint fits using pairs of tracers, i.e. \lya{} auto + \lya{} $\times$ QSO, \lya{} auto + \lya{} $\times$ GalAbs and \lya{} $\times$ QSO + \lya{} $\times$ GalAbs. We see that the combination \lya{}+GalAbs has $\sim20\%$ larger error for $\alpha_\parallel$ than \lya{}+QSO but that has $\sim10\%$ smaller errors compared to the combination QSO+GalAbs. Similarly, for $\alpha_\perp$ we see that the combination \lya{}+GalAbs has $\sim25\%$ larger errors than \lya{}+QSO and $\sim5\%$ smaller errors than the combination QSO+GalAbs. 

Finally, if we perform a joint fit using all the available correlations we find an overall decrease of the error bars (compared to the combination of \lya{}+QSO) of $\sim8\%$ for both $\alpha_\parallel$ and $\sim7\%$ for $\alpha_\perp$. Naturally, this need to be taken with a grain of salt as the \lya{} $\times$ SBLA cross-correlation has not been properly tested against systematics. Once again, we stress that the metal contamination model is not properly explaining the features observed in the new dataset. A refined set of mocks including the SBLAs and their associated metal contributions are required in order to clarify these points, but we leave this for the future.

\begin{table*}
    \centering
\scalebox{0.75}{
\begin{tabular}{l|lll|lll|l}
\toprule
    Parameters &              \lya{} &                 QSO & GalAbs &          \lya{}+QSO &     \lya{}+GalAbs &        QSO+GalAbs & \lya{}+QSO+GalAbs \\
    \midrule
    $\alpha_{\parallel}$ & $1.050\pm0.038$ & $1.044\pm0.042$ &        $1.000\pm0.080$ &    $1.044\pm0.027$ &            $1.031\pm0.033$ &            $1.036\pm0.036$ &                $1.034\pm0.025$ \\
    $\alpha_{\perp}$ & $0.974\pm0.046$ & $0.946\pm0.043$ &        $1.007\pm0.086$ &    $0.959\pm0.030$ &            $0.987\pm0.038$ &            $0.961\pm0.037$ &                $0.971\pm0.028$ \\
    $b_{\eta,{\rm Ly\alpha}}$ &   $-0.210\pm0.005$ &   $-0.210\pm0.007$ &   $-0.183\pm0.010$ &   $-0.193\pm0.010$ &   $-0.208\pm0.005$ &   $-0.184\pm0.010$ &    $-0.207\pm0.005$ \\
    $\beta_{\rm Ly\alpha}$ &      $2.03\pm0.09$ &      $2.03\pm0.10$ &      $2.05\pm0.09$ &      $2.03\pm0.09$ &      $2.03\pm0.09$ &      $2.05\pm0.09$ &       $2.02\pm0.09$ \\
    $b_{\eta,{\rm QSO}}$ &                    &             $1.00$ &                    &             $1.00$ &                    &             $1.00$ &              $1.00$ \\
    $\beta_{\rm QSO}$ &                    &    $0.244\pm0.012$ &                    &    $0.261\pm0.016$ &                    &    $0.204\pm0.017$ &     $0.252\pm0.015$ \\
    $\Delta r_{\parallel,{\rm QSO}} \left({\rm h^{-1}Mpc}\right)$ &                    &   $-0.240\pm0.621$ &                    &   $-0.291\pm0.628$ &                    &   $-0.279\pm0.619$ &    $-0.644\pm0.709$ \\
    $\sigma_{v,{\rm QSO}} \left({\rm h^{-1}Mpc}\right)$ &                    &    $8.941\pm2.413$ &                    &    $6.565\pm2.470$ &                    &    $9.186\pm2.275$ &     $7.718\pm2.252$ \\
    $b_{\eta,{\rm gal\,abs}}$ &                    &                    &              $1.0$ &                    &              $1.0$ &              $1.0$ &               $1.0$ \\
    $\beta_{\rm gal\,abs}$ &                    &                    &    $0.416\pm0.010$ &                    &    $0.424\pm0.010$ &    $0.416\pm0.010$ &     $0.423\pm0.010$ \\
    $\Delta r_{\parallel,{\rm gal\,abs}} \left({\rm h^{-1}Mpc}\right)$ &                    &                    &        $0.3\pm0.7$ &                    &       $-0.2\pm0.7$ &        $0.3\pm0.7$ &         $0.0\pm0.7$ \\
    $\sigma_{v,{\rm gal\,abs}} \left({\rm h^{-1}Mpc}\right)$ &                    &                    &        $0.0\pm7.6$ &                    &        $0.0\pm7.5$ &       $0.0\pm13.8$ &        $0.0\pm14.5$ \\
    $10^{3}b_{\eta,{\rm SiII(126)}}$ &           $-4\pm3$ &           $-4\pm4$ &          $-11\pm5$ &           $-3\pm2$ &           $-3\pm3$ &           $-6\pm3$ &            $-2\pm2$ \\
    $10^{3}b_{\eta,{\rm SiIII(120.7)}}$ &           $-7\pm3$ &           $-0\pm3$ &           $-0\pm2$ &          $-0\pm13$ &           $-5\pm3$ &           $-0\pm1$ &            $-3\pm3$ \\
    $10^{3}b_{\eta,{\rm SiII(119.3)}}$ &           $-6\pm3$ &           $-0\pm3$ &           $-0\pm2$ &           $-2\pm3$ &           $-2\pm3$ &           $-0\pm1$ &            $-0\pm3$ \\
    $10^{3}b_{\eta,{\rm SiII(119)}}$ &          $-0\pm15$ &           $-0\pm3$ &           $-0\pm3$ &           $-0\pm3$ &           $-0\pm3$ &           $-0\pm1$ &            $-0\pm2$ \\
    $10^{3}b_{\eta,{\rm CIV({\rm eff})}}$ & $-0.0051\pm0.0026$ & $-0.0050\pm0.0026$ & $-0.0050\pm0.0025$ & $-0.0051\pm0.0026$ & $-0.0056\pm0.0027$ & $-0.0050\pm0.0025$ &  $-0.0056\pm0.0027$ \\
    $b_{\rm HCD}$ &   $-0.000\pm0.145$ &           $-0.000$ &           $-0.000$ &   $-0.033\pm0.016$ &   $-0.000\pm0.014$ &   $-0.000\pm0.006$ &    $-0.000\pm0.132$ \\
    $\beta_{\rm HCD}$ &    $0.500\pm0.090$ &            $0.500$ &            $0.500$ &    $0.499\pm0.090$ &    $0.500\pm0.089$ &    $0.500\pm0.089$ &     $0.500\pm0.089$ \\
    $10^{2}A_{\rm sky, Ly\alpha(Ly\beta)}$ &    $0.006\pm0.001$ &                    &                    &    $0.006\pm0.001$ &    $0.006\pm0.001$ &                    &     $0.006\pm0.001$ \\
    $\sigma_{\rm sky, Ly\alpha(Ly\beta)}$ &       $33.8\pm3.2$ &                    &                    &       $33.9\pm3.2$ &       $33.9\pm3.1$ &                    &        $34.0\pm3.2$ \\
    \midrule
    $\chi^{2}$ &           2440.543 &           5005.963 &          10164.458 &           7449.689 &          12619.806 &          15172.998 &           17628.191 \\
    $N_{\rm bin}$ &               2470 &               4940 &               9880 &               7410 &              12350 &              14820 &               17290 \\
    $N_{\rm param}$ &                 15 &                 12 &                 36 &                 18 &                 42 &                 41 &                  45 \\
    probability &              0.578 &              0.215 &              0.012 &              0.316 &              0.024 &              0.011 &               0.020 \\
    \midrule
    $b_{\rm Ly\alpha}$ &  $-0.101\pm-0.005$ &  $-0.100\pm-0.005$ &  $-0.086\pm-0.004$ &  $-0.092\pm-0.004$ &  $-0.099\pm-0.005$ &  $-0.087\pm-0.004$ &   $-0.100\pm-0.005$ \\
    $b_{\rm QSO}$ &                    &    $3.978\pm0.196$ &                    &    $3.718\pm0.229$ &                    &    $4.755\pm0.398$ &     $3.857\pm0.236$ \\
    $b_{\rm gal\,abs}$ &                    &                    &    $2.333\pm0.057$ &                    &    $2.291\pm0.052$ &    $2.333\pm0.057$ &     $2.292\pm0.052$ \\
\bottomrule
\end{tabular}
}
    \caption{Best-fit parameters for different combinations of tracers. All the fits are shown using the \lya{} flux transmission field with masked SBLAs. Values without error bars were fixed to the given values and empty values indicate the parameters that were not used in some of the fits. Error bars are given by \texttt{iminuit} for all parameters except for $\alpha_{\parallel}$ and $\alpha_{\perp}$. These two parameter errors are estimated from a $\chi^2$ scan. The first block contains fits using only one set of tracers (\lya{} for the \lya{} auto-correlation and QSO for the \lya{} $\times$ QSO cross-correlation). The second block contains fits using pairs of tracers and the final block contains fits using all available tracers. Here, {\it GalAbs} stands for the cross-correlation between the \lya{} forest and galaxies in absorption, i.e. the combination of \lya{} $\times$ DLAs and \lya{} $\times$ SBLAs. The bias for the different tracers is derived at the bottom of the table (but not used in the fit). Note that each of the fits uses the correlations from the \lya{} and \lyb{} regions.}
    \label{tab:joint_fit_param}
\end{table*}
\section{Discussion}\label{sec:discussion}
\subsection{Importance of masking SBLAs. Implications of the measured biases}
In section~\ref{sec:cross_covariance}, we have seen that masking SBLAs in the computation of the \lya{} flux transmission field helps in reducing the cross-covariance between the samples. This is convenient because it simplifies a joint fit of the different correlation functions. What is more, the constraints on BAO parameters are minimally affected by the extra masking (see section~\ref{sec:masking_effects}).

However, the importance of masking SBLAs in the computation of the \lya{} flux transmission field goes beyond this. \lya{} absorption within the \lya{} forest is known to come from two distinct sources. On the one hand, we have the intergalactic medium (IGM), which is diffuse and only mildly non-linear and hence has a small clustering bias. On the other hand, \lya{} and associated metal absorption arising from the circumgalactic medium (CGM), are associated with collapsed structures and therefore a larger bias is expected.

Using an inhomogeneous \lya{} forest tracer leads to certain challenges and constraints. The fitted tracer bias and redshift space distortions become hybrid quantities and the relative contributions of the underlying source clustering properties are difficult to disentangle. Furthermore, metals are highly inhomogeneous and largely associated with the CGM.  Metal fits (and metals in mocks derived from these fits) that neglect this is unrepresentative of the true population, potentially concealing important effects embedded in correlation function estimates.
In this context, the measurement of \lya{} forest tracer bias with and without the masking of SBLAs is potentially informative since it could indicate whether a more homogeneous tracer has indeed been obtained.

Having SBLAs removed from the \lya{} forest implies that the degree of mixing of the \lya{} forest is now lower. In a simplified picture, one could argue that there are only two phases to the \lya{} forest and that now we can interpret the \lya{} bias  as a physical quantity. This would imply that the IGM clustering is weaker than that of the CGM. This could mean that the IGM is found between large halos, connecting them, but a more detailed study is required to understand this. While this picture does make sense, it is important to note that we have only made a first step towards the full understanding of the \lya{} forest and a more detailed study of the different phases of the forest is required before any solid claims are made. 

\subsection{Implications for cosmological mocks}
Synthetic spectra, or mocks, are used to test the models of the \lya{} correlation functions \citep[e.g.][]{Farr+2020, Etourneau2020}. However, the results of this work indicate that the current \lya{} mocks are not sufficiently complex for a complete description of possible systematic effects. For instance, the population of galaxies in absorption is incomplete in these mocks given that they include only High Column Density Absorbers (including DLAs and sub-DLAs, but with column densities much higher than those typically measured in SBLAs), which also do not have associated metals).

Having a realistic population of galaxies in absorption present in the mocks is desirable for a full description of the correlation function including physically motivated broadband functions. Having more reliable mocks is necessary to correctly measure Alcock-Paczynski and Redshift Space Distortions \citep[see e.g][]{Cuceu+2021}, as well as reliable metals and a meaningful \lya{} bias.  For instance, masking SBLAs seems to have an important impact on the measured \lya{} and quasar bias parameters (see section~\ref{sec:masking_effects}). We need these objects in the mocks in order to properly test these changes.

What is more, from stacking results such as those from \cite{Pieri+2010, Pieri+2014, Yang+2022} or preferably full population analysis such as those from \cite{Morrison+Inprep}, we see that SBLAs have a high level of associated metal absorption. This metal absorption has a strong impact on the correlation function at small perpendicular separations, the scales where our fit model fails. In the current, LyaCoLoRE mocks \citep{Farr+2020}, metals are ``painted" on following the \lya{} forest  opacity with only coefficients determining their relative strengths derived from metal model fits to the 3D correlation function. High levels of astrophysical realism were not the goal. The main purpose was to develop physical models that are sufficiently sophisticated to fit the eBOSS DR16 \lya{} auto- and quasar cross-correlations.  This approach provides a simple means to test the continuity between metal addition to mocks and their fits. In fact, the metal fit in the correlation function is somewhat more sophisticated than the prescription for metal addition derived from it; the redshift space distortion for the metals is fixed $\beta = 0.5$ \citep{duMasdesBourboux+2020}, which is a reasonable value for galaxies in absorption as opposed to the \lya{} forest redshift space distortions used in LyaCoLoRE mocks. 

The new measurements presented here suggest that the metal model lacks realism even for the simplistic treatment of metals as contaminants to the 3D correlation function when a wider set of correlation functions are considered. To use the new correlation functions presented here (to both boost statistical constraints and better separate the forest into diffuse/collapsed components) a more realistic metal mock prescription and subsequent metal model fit must be explored. 

More sophisticated metal distributions have been added to previous versions of \lya{} forest mocks \citep{Slosar+2011,Bautista+2015}. They introduced non-linear relationships between \lya{} and metal opacities, and a degree of scattering in the metal strength to generate various effects in mean composite spectra, but the metal strength distribution was unrefined and the metals made no reference to realistic galaxy positions. SBLA bias measurements (here and in \citealt{Morrison+Inprep}) along with metal population analysis presented in \cite{Morrison+Inprep}, now provide the observational input for realistic metal distributions in mocks following the galaxies in absorption that dominate large \lya{} forest survey datasets.

\subsection{Impact on upcoming surveys}
The work in this paper uses data from the eBOSS survey, but it will have more impact on the next generation of surveys over the coming years. The work presented here will allow for fuller exploitation of the data and enable analysis methods (e.g. cosmological mocks and correlation function modelling) to be sufficiently refined when confronted with next-generation sample precision.
Two next-generation surveys show the clearest potential for measuring the clustering of galaxies in absorption with the \lya{} forest. Firstly, the DESI survey \citep{Desi+2016a, Desi+2016b} will observe an unparalleled large sample of \lya{} quasars, where SBLAs are detected. This survey has already started operations gathering quasar spectra \citealt{Alexander+2022}). 
DESI will be taking spectra of $\sim50$ \lya{} quasars per square degree, a factor of three higher than the 17 \lya{} quasars per square degree observed in BOSS+eBOSS \citep{Desi+2016a}. DESI will observe around 700 000 \lya{} forest quasar spectra.

Secondly, the WEAVE-QSO survey \citep{Pieri+2016} as part of the wider WEAVE survey \citep{Dalton+2016} will observe fewer \lya{} quasar spectra than DESI but with higher density at $z>2.5$ and higher spectral resolution (\citealt{Pieri+2016}, Pieri et al. in prep). The first light with the WEAVE spectrograph on the William Herschel Telescope is expected by the end of 2022. WEAVE-QSO will observe more than 400 000 \lya{} forest quasar spectra.

In addition to the larger samples, these next-generation surveys will provide higher signal-to-noise allowing tests for SBLA purity and completeness will be improved for the same quasars.  DESI \lya{} will have approximately twice the spectral resolution and  WEAVE-QSO will have three times the resolution compared to the (e)BOSS spectra. This increased resolution opens the possibility to assess smaller blending scales than that of the current samples of SBLAs, which is limited to that of the (e)BOSS line spread function. WEAVE-QSO and DESI will offer the potential to recover a broader population of galaxies in absorption, through the freedom to treat the absorption grouping scale as a free parameter.

\section{Summary and conclusions}\label{sec:conclusions}
We have presented the first measurements of the \lya{} $\times$ SBLA cross-correlations. We compared these correlations with the \lya{} $\times$ DLA cross-correlation and found that they have similar clustering strengths, suggesting that they come from the same type of systems, namely galaxies in absorption. We also presented the first BAO measurement of these high-z ($z>2$) galaxies and compared its constraining power with that of the \lya{} auto-correlation and the \lya{} $\times$ QSO cross-correlation. Our main conclusions are as follows:
\begin{itemize}
    \item Our clustering measurements suggest that DLAs and SBLAs reside in the same population of halos.
    \item The removal of SBLA from the \lya{} forest allows for a cleaner interpretation of the \lya{} forest bias factor, as we make progress towards separating the contribution of CGM absorption from the overall IGM forest signal. This is also an analysis requirement since the use of SBLAs as a separate tracer requires their masking in the forest tracer to diminish cross-covariance between the \lya{} $\times$ SBLA  correlation function and standard correlation functions to acceptable levels.
    \item The BAO scale uncertainty from the \lya{} $\times$ SBLA cross-correlation are 1.75$\times$ that of the \lya{} auto-correlation and 1.6$\times$ the \lya{} $\times$ QSO cross-correlation if the tracers are used on their own. By combining all three probes we can decrease the error bars by $\sim8\%$ for $\alpha_\parallel$ and $\sim7\%$ for $\alpha_\perp$ compared to the most up-to-date measurements from eBOSS \citep{duMasdesBourboux+2020}.
    \item Tests with mocks are needed in order to build confidence in the BAO measures explored here. These combined results should be taken as a demonstration and not used to constrain cosmological models.
    \item 
    We find that the correlation function fit model breaks at small perpendicular separations ($r_{\perp} < 30$\hmpc{}) for the correlation \lya{} $\times$ SBLA. We excluded these points from our fit, but the model should be revised and tested on mocks once they become available. The cause for this breakdown is likely to be the metal fit portion of this model since SBLAs are the sites of particularly strong metals, and such sites are not taken into account in the fitting.
    \item The already running DESI survey and the upcoming WEAVE-QSO survey will boost the precision of this measurement.
    Higher confidence in SBLA detection and greater flexibility to broaden the selection of strong blended \lya{} absorption to different blending scales in this new data will  make the measurement more robust.
\end{itemize}

\section*{Acknowledgements}
The authors would like to thank Andreu Font-Ribera and James Rich for very useful discussions and comments. 

This project has received funding from the European Union’s Horizon 2020 research and innovation program under the Marie Skłodowska-Curie grant agreement No. 754510.

This work was supported by the A*MIDEX project (ANR-11-IDEX-0001-02) funded by the ``Investissements d'Avenir'' French Government program, managed by the French National Research Agency (ANR), by ANR under contract ANR-14-ACHN-0021 and by the Programme National Cosmology et Galaxies (PNCG) of CNRS/INSU with INP and IN2P3, co-funded by CEA and CNES.

AC acknowledges support from the United States Department of Energy, Office of High Energy Physics under Award Number DE-SC-0011726.

IFAE is partially funded by the CERCA program of the Generalitat de Catalunya.

Funding for the Sloan Digital Sky Survey III/IV has been provided
by the Alfred P. Sloan Foundation, the U.S. Department of Energy
Office of Science, and the Participating Institutions. SDSS-III/IV
acknowledge support and resources from the Center for High
Performance Computing at the University of Utah. The SDSS
The website is \url{www.sdss.org}. SDSS is managed by the Astrophysical Research Consortium for the Participating Institutions of the SDSS Collaboration including the Brazilian Participation Group, the Carnegie Institution for Science, Carnegie Mellon University, Center for Astrophysics | Harvard \& Smithsonian, the Chilean Participation Group, the French Participation Group, Instituto de Astrofísica de Canarias, The Johns Hopkins University, Kavli Institute for the Physics and Mathematics of the Universe (IPMU) / University of Tokyo, the Korean Participation Group, Lawrence Berkeley National Laboratory, Leibniz Institut für Astrophysik Potsdam (AIP), Max-Planck-Institut für Astronomie (MPIA Heidelberg), Max-Planck-Institut für Astrophysik (MPA Garching), Max-Planck-Institut für Extraterrestrische Physik (MPE), National Astronomical Observatories of China, New Mexico State University, New York University, University of Notre Dame, Observatário Nacional / MCTI, The Ohio State University, Pennsylvania State University, Shanghai Astronomical Observatory, United Kingdom Participation Group, Universidad Nacional Autónoma de México, University of Arizona, University of Colorado Boulder, University of Oxford, University of Portsmouth, University of Utah, University of Virginia, University of Washington, University of Wisconsin, Vanderbilt University, and Yale University.

\section*{Data availability}
The catalogue of SBLAs will be made public upon publication. We refer the reader to \cite{duMasdesBourboux+2020} and \cite{Chabanier+2021} for the data access to the SDSS spectra used to generate \lya{} flux transmission field and the DLA catalogue respectively. The software used in this analysis are publicly available on GitHub: \texttt{picca} \url{https://github.com/igmhub/picca/releases/tag/v4}; and \texttt{vega} \url{https://github.com/andreicuceu/vega/tree/master/vega}.

\bibliographystyle{apj}
\bibliography{main}

\appendix
\section{\lya{}(\lyb{}) correlation functions}\label{sec:lya(lyb)}
\begin{figure}
    \centering
    \includegraphics[width=\columnwidth]{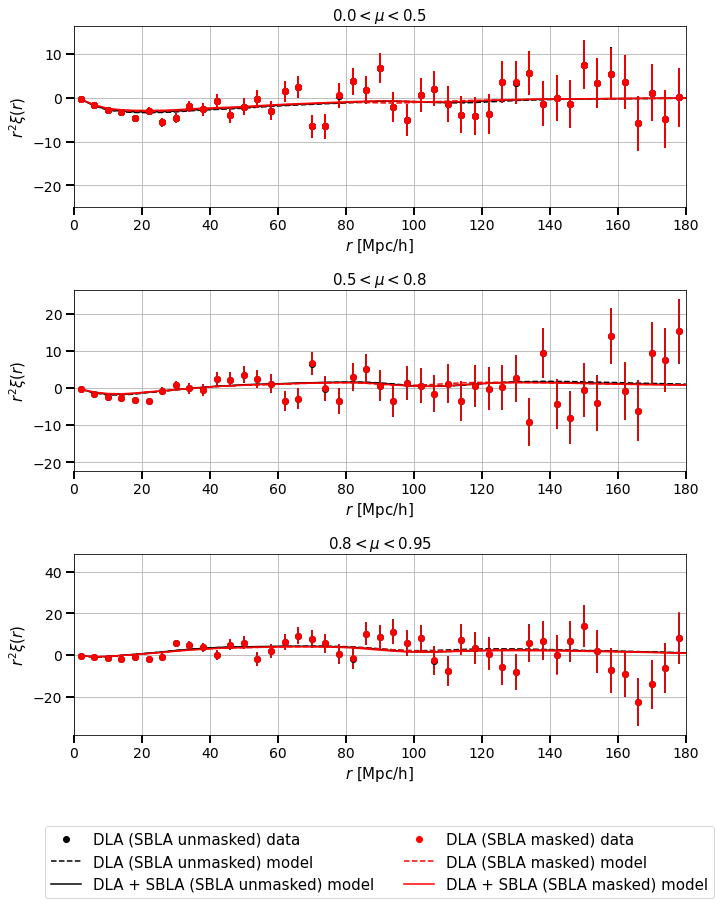}
    \caption{Same as figure~\ref{fig:dlaxlya} but for \lya{}(\lyb{}) $\times$ DLAs. We see that the correlation function remains unchanged when SBLAs are masked from the computation of the \lya{} flux transmission field. }
    \label{fig:dlaxlyb}
\end{figure}

\begin{figure}
    \centering
    \includegraphics[width=\columnwidth]{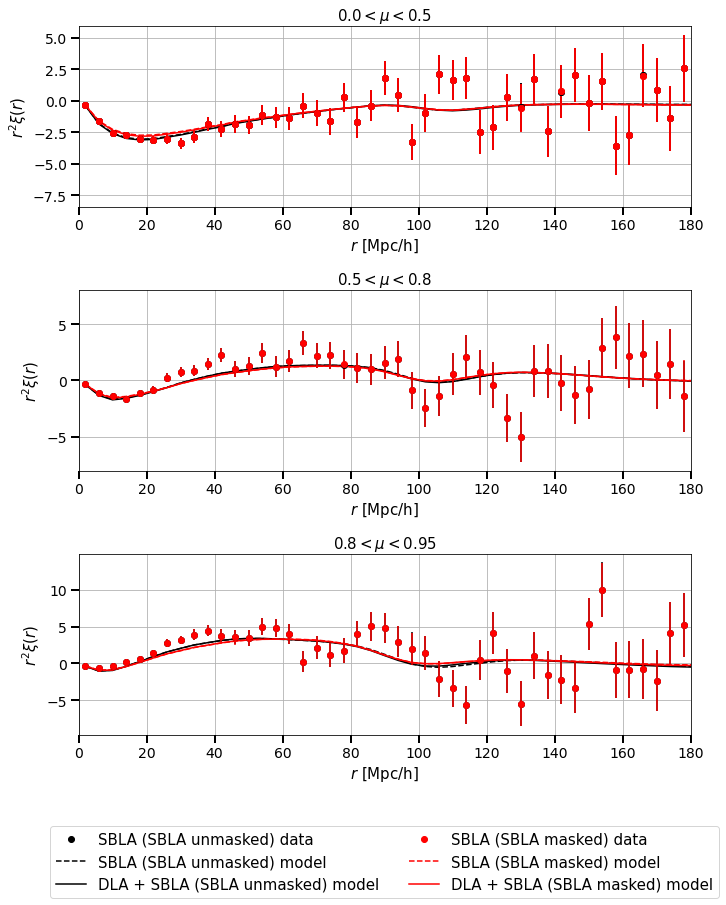}
    \caption{Same as figure~\ref{fig:dlaxlya} but for \lya{}(\lyb{}) $\times$ DLAs. We see that the correlation function remains unchanged when SBLAs are masked from the computation of the \lya{} flux transmission field. }
    \label{fig:sblaxlyb}
\end{figure}

In section~\ref{sec:results} we have shown the cross-correlation of \lya{}(\lya{}) $\times$ DLAs and \lya{}(\lya{}) $\times$ SBLAs and mentioned that the reported fits also included the correlations with \lya{}(\lyb{}). For completeness, here we show the equivalent plots for the \lya{}(\lyb{}) cross-correlations. Figure~\ref{fig:dlaxlyb} contains the \lya{}(\lyb{}) $\times$ DLA cross-correlation and figure~\ref{fig:sblaxlyb} contains the \lya{}(\lyb{}) $\times$ SBLA cross-correlation. We note that there are no changes to these correlation functions when masking SBLAs from the computation of the \lya{} flux transmission field since there are no SBLAs in the \lya{}(\lyb{}) field and thus they remain unchanged.

\end{document}